\documentstyle[12pt,aaspp4,flushrt]{article}
\slugcomment{Astrophysical Journal in press}

\lefthead{Alves et al.}
\righthead{Correlation Between Gas and Dust}

\def\h2{H$_2$\ }

\def\pc3{pc$^{-3}$}

\def\h {$^h$}

\def\deg{$^{\circ}$}

\def\as{$^{\prime\prime}$}
\def\ltsima{$\;\buildrel<\over\sim\;$} 
\def\simlt{\lower.5ex\hbox{\ltsima}}
\def\gtsima{$\;\buildrel>\over\sim\;$}
\def\simgt{\lower.5ex\hbox{\gtsima}}
\def\cao{\c{c}\~{a}o}

\begin{document}

\title{Correlation Between Gas and Dust in Molecular Clouds: L977}

\vskip 0.2cm
\author{Jo\~ao  Alves\altaffilmark{1,2}, Charles J. Lada
\altaffilmark{1}}
\affil{Harvard-Smithsonian Center for Astrophysics, 
60 Garden St., Cambridge MA 02138}
\author{Elizabeth A. Lada\altaffilmark{1}}
\affil{Astronomy Department, University of Florida,
Gainesville, FL 32608}

\authoremail{jalves@cfa.harvard.edu}

\altaffiltext{1}{Visiting Astronomer, Kitt National Observatory, part of the
National Optical Astronomy Observatories, which is operated by the Association
of Universities for Research in Astronomy, Inc. under contract with the
National Science Foundation.}
\altaffiltext{2}{Also Physics Department, University of Lisbon, Lisbon Portugal}

\begin{abstract}

We report observations of the $J =$ (1--0) C$^{18}$O molecular emission
line toward the L977 molecular cloud.  To study the correlation between
C$^{18}$O emission and dust extinction we constructed a Gaussian smoothed
map of the infrared extinction measured by Alves et al.  (1998) at the
same angular resolution (50\as) as our molecular--line observations.  This
enabled a direct comparison of C$^{18}$O integrated intensities and column
densities with dust extinction over a relatively large range of cloud
depth (2 $<$ A$_V$ $<$ 30 mag) at 240 positions inside L977.  We find a
good linear correlation between these two column density tracers for cloud
depths corresponding to A$_V$ $\la$ 10 magnitudes.  For cloud depths
above this threshold there is a notable break in the linear correlation. 
Although either optically thick C$^{18}$O emission or extremely low
($T_{ex} <$ 5 K) excitation temperatures at high extinctions could produce
this departure from linearity, CO depletion in the denser, coldest regions
of L977 may be the most likely cause of the break in the observed
correlation.  We directly derive the C$^{18}$O abundance in this cloud
over a broad range of cloud depths and find it to be virtually the same as
that derived for IC 5146 from the data of Lada et al.  (1994). 

In regions of very high extinction (A$_V > $ 10 mag), such as dense cores,
our results suggest that C$^{18}$O would be a very poor tracer of mass.
Consequently, using C$^{18}$O as a column density tracer in molecular
clouds can lead to a 10 to 30\% underestimation of overall cloud mass. 

We estimate the minimum total column density required to shield C$^{18}$O
from the interstellar radiation field to be 1.6 $\pm$ 0.5 magnitudes of
visual extinction.

\end{abstract}


\newpage

\section{Introduction}

Molecular clouds are the coolest (T $\sim$ 10 K) objects in the known
Universe.  They contain about half of the mass of the Interstellar Medium
(ISM) and hence an important fraction of the mass of the Galaxy.  Yet, the
most important characteristic of molecular clouds is that they are the
sites of all galactic star formation.  To understand how stars are formed
one needs to understand how molecular clouds evolve.  A prerequisite to
this understanding is the detailed knowledge of the physical and chemical
structure of molecular clouds. 

About 99\% of the mass of a molecular cloud is composed of hydrogen and
helium making such clouds virtually inaccessible to direct observation. 
For this reason, the traditional method used to derive the basic physical
properties of these objects (e.g., sizes, masses, temperatures) is the
spectroscopic observation of rare but detectable trace molecules (CO, CS,
NH$_3$) which are several orders of magnitude ($\sim$ 4--9) less abundant
than the primary mass component, molecular hydrogen (H$_2$).  However, the
interpretation of such molecular--line observations is not always
straightforward.  Several poorly constrained effects (e.g., deviations
from local thermodynamic equilibrium, opacity variations, chemical
evolution, small scale cloud structure, depletion) compromise the
derivation of such important properties such as the distribution of mass
and structure of a molecular cloud.  A better quantitative understanding
of the relationship of these tracers to H$_2$ is necessary not only for
the derivation of reliable masses but also for knowledge of the chemical
structure of molecular clouds. 

Because of the apparent constancy of the gas--to--dust ratio in
interstellar clouds (e.g., Lilley 1955; Jenkins \& Savage 1974; Bohlin et
al.  1978) the most reliable way to trace the total gas content of a
molecular cloud may be to measure the distribution of dust through it. 
The recent development of sensitive near-infrared array cameras has
enabled measurements of the dust distribution in certain molecular clouds
over an unprecedented range of depth and angular scale (Lada et al.  1994;
Alves et al.  1998).  With current detectors and small telescopes (1--2m)
we can readily detect reddened field stars behind a molecular cloud with
as much as 25--30 magnitudes of visual extinction, approximately an order
of magnitude improvement over previous measurements.  This allows the
mapping of the distribution of dust at an angular resolution and optical
depth that are suitable to probe directly the inner, denser regions of a
molecular cloud where star formation takes place.  Combined with molecular
spectroscopy, such observations open a new window on molecular cloud
research enabling direct measurements of molecular abundances to cloud
depths not previously accessible.  Such observations will have important
consequences for understanding molecular cloud chemistry.  For example,
Lada et al.  (1994) in their study of the IC 5146 molecular cloud found
that the ratio $^{13}$CO/C$^{18}$O appears to be a strong function of
extinction --- suggesting that one or both CO species are unstable in the
outer regions (A$_V$ $<$ 10 mag) of the cloud --- probably due to active
chemical processing by UV radiation.  Their results also suggested the
presence of C$^{18}$O depletion in the densest regions of the IC 5146
cloud which was recently confirmed by higher resolution C$^{18}$O and
C$^{17}$O observations by Kramer et al.  (1998) obtained with the IRAM 30m
telescope.

Recently, Alves et al.  (1998) used measurements of near-infrared color
excess of $\sim$ 2000 stars to directly measure and map dust extinction
through the L977 molecular cloud.  L977 is a well defined dark cloud lying
against a rich background of field stars towards the warped plane of the
Galaxy ($l = 90$\deg, $b = 2$\deg).  It was mapped in $^{13}$CO, with
2$^\prime$.7 resolution, as part of a large scale survey of the region
(Dobashi et al.  1994).  The $^{13}$CO survey showed that L977 is a
relatively isolated molecular cloud, free from confusion with other clouds
along the line--of--sight.  Alves et al. (1998) were also able to derive a
distance, $d$, of 500 pc towards L977 and a mass of $M_{L977} = (660 \ \pm
\ 30) (d/500 pc)^2$ M$_\odot$. 

The main goal of this article is to study the variation of C$^{18}$O
molecular abundance with dust extinction (e.g., A$_V$) over a large range
of cloud depth (2 $<$ A$_V$ $<$ 30 mag).  C$^{18}$O is often optically
thin in molecular clouds and widely used as a tracer of molecular
hydrogen, H$_2$.  We perform a direct comparison of C$^{18}$O integrated
intensity and column density with measurements of dust extinction with the
same angular resolution (50\as), for 240 positions in molecular cloud
L977.

We describe the acquisition and reduction of the observations in Section 2
of the paper.  In Sec.  3 we present the results of the observations and
in Sec.  4 we present the analysis and discussion. In Sec.  5 we summarize
our conclusions.

\section{Observations and Data Reduction}

We employed the 14m Five College Radio Astronomy Observatory (FCRAO)
telescope in New Salem, MA, to obtain observations of the 109 GHz $J = $
1--0 C$^{18}$O molecular emission line toward L977.  Observations at FCRAO
\footnote{The Five College Radio Astronomy Observatory is operated with
the support of the National Science Foundation, under grant AST 94-20159
and with permission of the Metropolitan District Commission of
Massachusetts.} were made in November 1995 and February 1996 utilizing the
QUARRY 15 element focal plane array (Erickson et al.  1992) and the FAAS
autocorrelator to generate a single beam-sampled map ($\sim$ 50\as\
resolution) covering an area on the sky of 10$\times$15 arcmin centered on
the L977 molecular cloud.  We used the 40 MHz bandwidth that yielded a
channel spacing of 78 kHz and a velocity resolution of 0.215 kms$^{-1}$ at
109 GHz.  The observations were frequency switched, with the reference
frequency displaced from the signal frequency by 4 MHz.  System
temperatures during observations ranged from 480 to 620 K.  Integration
times were adjusted so that the rms noise temperatures in the 78 kHz
channels was $\sim$ 0.13 K for the C$^{18}$O observations.  The half-power
beam width ($\Theta_{MB}$(109 GHz) $=$ 46.1\as) and beam efficiency
($\eta_B =$ 45\%) were taken from Ladd \& Heyer 1996.  All observations
were calibrated by a chopper wheel which allowed switching between the sky
and an ambient temperature load.  All data were reduced using the SPA and
CLASS spectral line reduction packages. The spectra were folded and linear
baselines were removed from the spectra.  An rms pointing uncertainty of
less than 5\as\ was determined from repeated observations of SiO maser
sources U Her and T Cep.

\section{Results}
\subsection{Molecular-Line Observations}

\begin{figure}
\plotfiddle{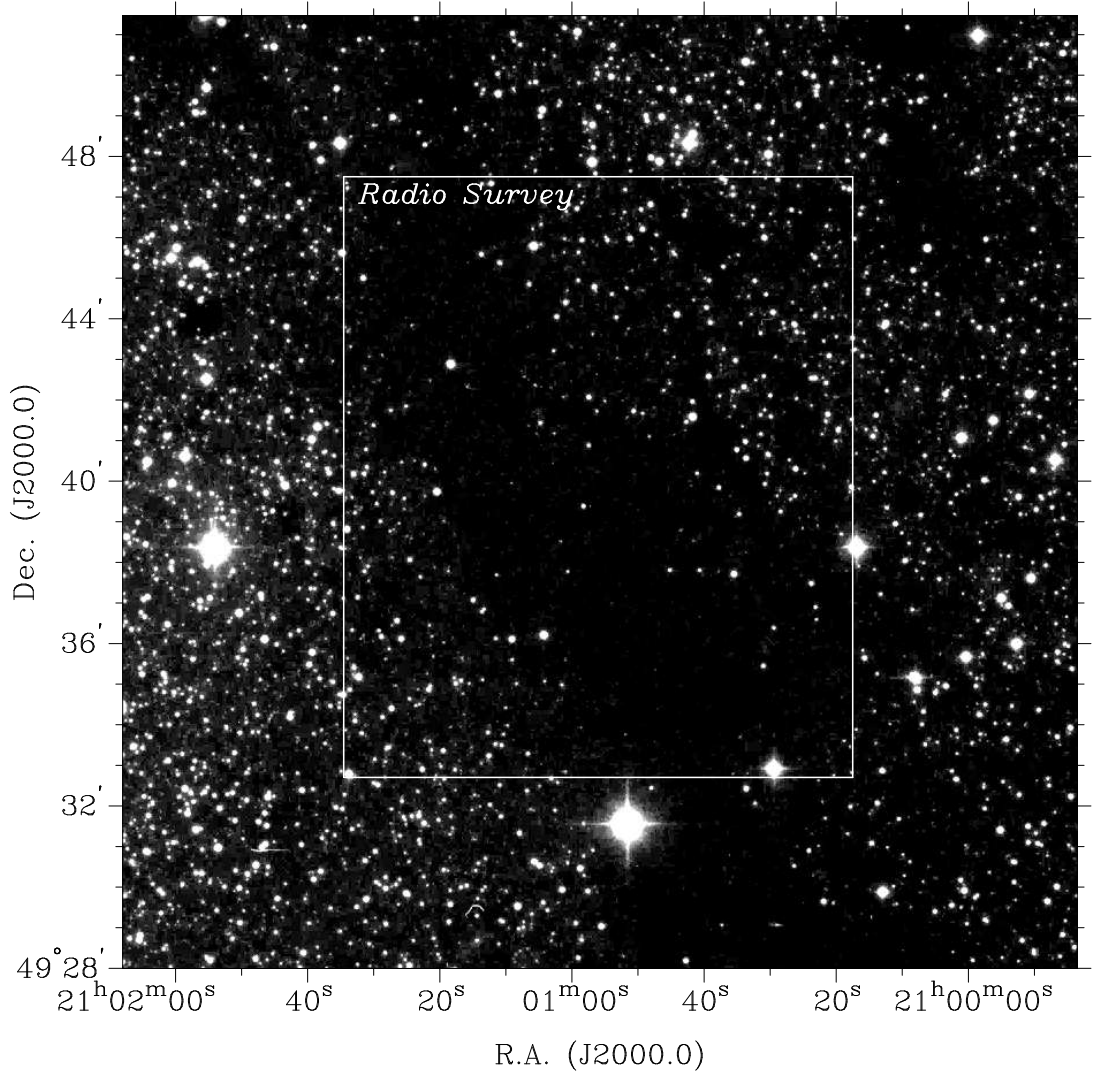}{14cm}{-90}{80}{80}{-310}{490}
\caption[Finding chart for L977 molecular cloud] 
{Finding chart for L977 molecular cloud.  The
area surveyed for C$^{18}$O molecular--line emission is represented as
the central rectangle overlaid on the Digitized Sky Survey red POSS
plate.  The molecular cloud is readily seen as the zone of obscuration
inside this square and seen against the rich star field that
characterizes this region of the galaxy.  \label{palomar}}
\end{figure}

\begin{figure}
\plotfiddle{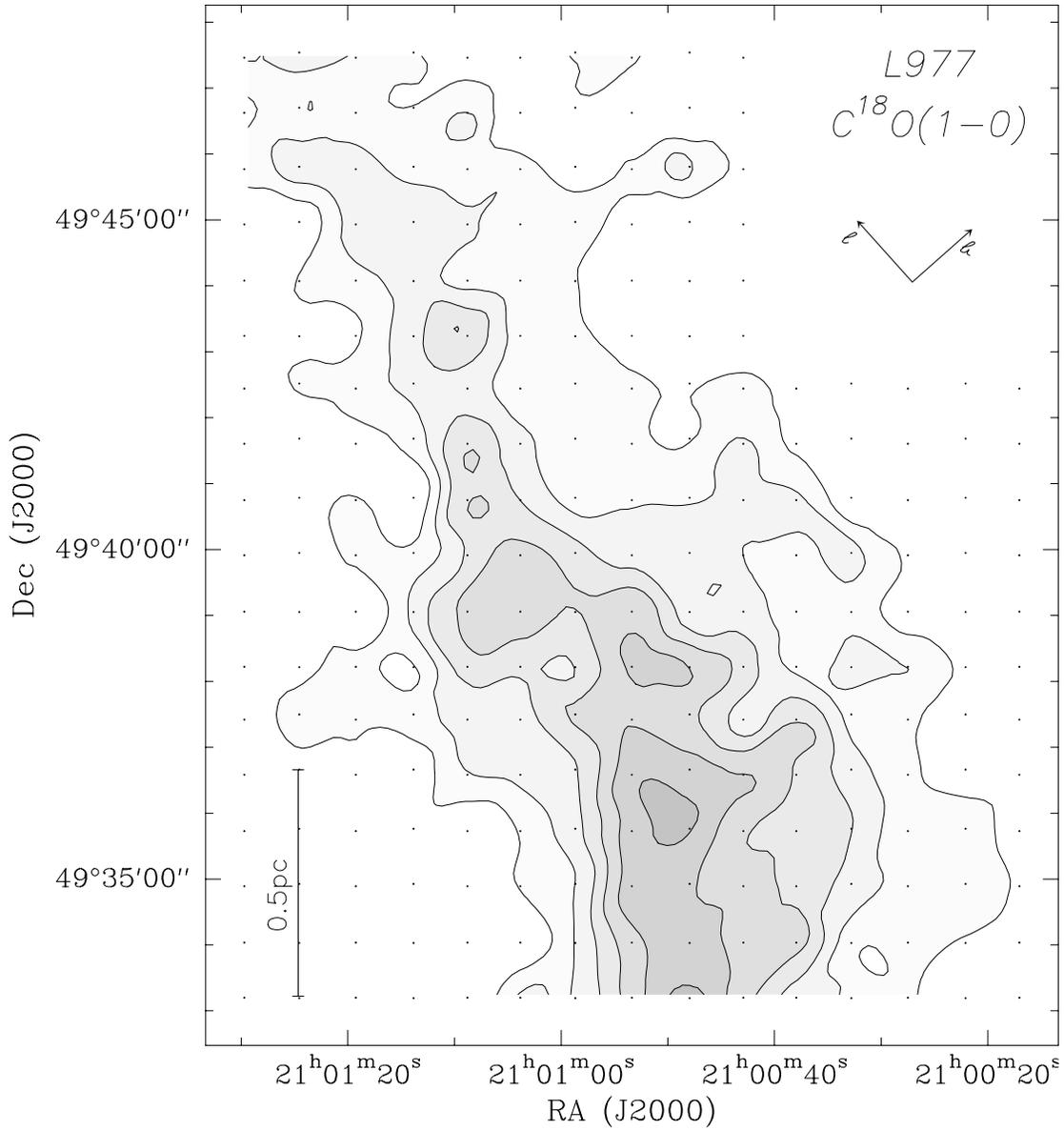}{14cm}{0}{75}{75}{-250}{-60}
\caption[Map of the integrated C$^{18}$O (1--0) emission
observed towards L977] {Map of the integrated C$^{18}$O (1--0) emission
observed towards L977 between $-2.0$ $<$ v$_{LSR}$ $<$ 0.0 kms$^{-1}$.
The contours start at 20\% (4 $\sigma$) and go up to 95\% in steps of
15\% (3 $\sigma$) of the peak integrated intensity (2.0 K kms$^{-1}$).
The
straight line at the lower left corner of the map represents 0.5 pc at
an assumed distance to L977 of 500 pc. \label{c18ototal}}
\end{figure}

The area surveyed for (1--0) C$^{18}$O molecular--line emission is
represented in Figure~\ref{palomar} as the central rectangle overlaid on
the Digitized Sky Survey red POSS plate \footnote{Based on photographic
data of the National Geographic Society -- Palomar Geographic Society to
the California Institute of Technology.  The plates were processed into
the present compressed digital form with their permission.  The Digitized
Sky Survey was produced at the Space Telescope Science Institute under US
Government grant NAG W-2166.}. The L977 molecular cloud is readily seen as
the zone of obscuration against the rich star field that characterizes
this region of the galaxy.  The map of the integrated C$^{18}$O (1--0)
emission observed towards L977 is given in Figure~\ref{c18ototal}.  The
intervals of integration ($-2.0$ $<$ v$_{LSR}$ $<$ 0.0 kms$^{-1}$) were
chosen from analysis of the average of all 240 observed positions to
include the region in the spectra where most of the signal is detected. 
The contours start at 20\% (4 $\sigma$) and go up to 95\% in steps of 15\%
(3 $\sigma$) of the peak integrated intensity (2.0 K kms$^{-1}$).  The
(0,0) point in this map corresponds to [$\alpha,\delta(2000.0) =
21^h00^m28^s, +49^{\circ}34^{\prime} 54^{\prime\prime}$].  The straight
line at the lower left corner of the map represents 0.5 pc at an assumed
distance to L977 of 500 pc.  The galactic coordinates reference frame is
shown at the upper left corner.  Comparison to the Palomar red plate
(Figure~\ref{palomar}) reveals a good correlation between C$^{18}$O
emission and the shape of the more opaque regions in L977.  The emission
follows closely the NE---SW elongated dark lane that characterizes this
molecular cloud at optical wavelengths.

\begin{figure}
\plotfiddle{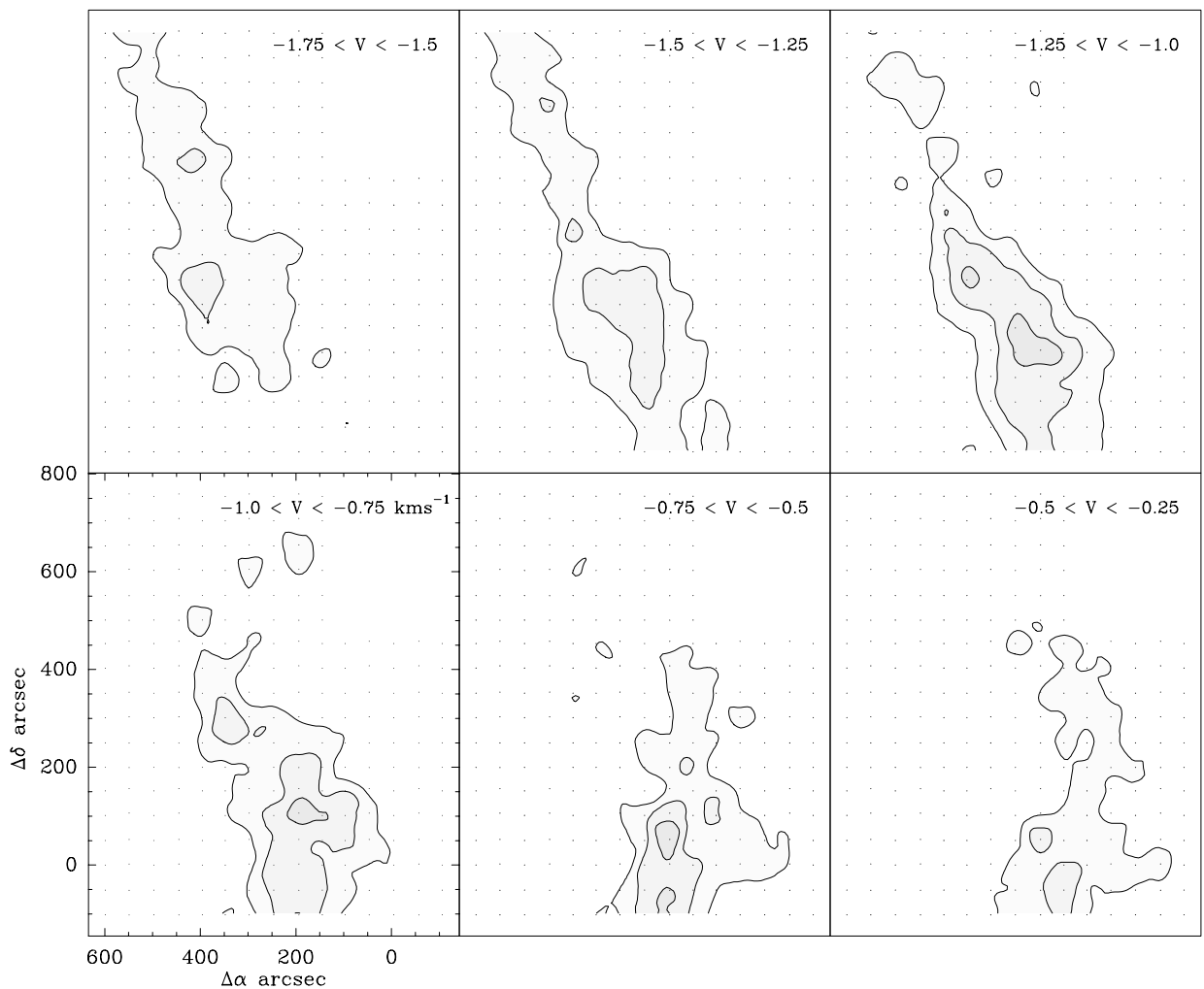}{14cm}{-90}{80}{80}{-310}{450}
\caption[L977 C$^{18}$O (1--0) kinematics] 
{L977 C$^{18}$O (1--0) kinematics.  Each of the
four maps were integrated every 0.25 kms$^{-1}$ from $-$1.75 to
$-$0.25 kms$^{-1}$.  Contours start at 4 $\sigma$ (0.13 K kms$^{-1}$)
and increase in steps of 4 $\sigma$.  
Most of the emission is found in
the range $-$2 to 0 kms$^{-1}$ with an average systemic
velocity of $-$1.0 kms$^{-1}$. 
The (0,0) point in this map corresponds to [$\alpha,\delta(2000.0) =
21^h00^m28^s, +49^{\circ}34^{\prime} 54^{\prime\prime}$].
\label{c18ovel}}
\end{figure}

The C$^{18}$O (1--0) kinematics are presented in Figure~\ref{c18ovel}.
Each of the six maps were integrated over 0.25 kms$^{-1}$ intervals from
$-$1.75 to $-$0.25 kms$^{-1}$ respectively.  Contours start at 4 $\sigma$
(0.13 K kms$^{-1}$) and increase in steps of 4 $\sigma$. Most of the
emission is found in the narrow range $-$2 to 0 kms$^{-1}$ with an average
systemic velocity of $-$1.0 kms$^{-1}$.  Although we are looking near a
complicated region of the galaxy there appears to be no confusion with
foreground/background molecular clouds.  L977 C$^{18}$O emission appears
narrowly confined in both space and velocity.  These results are
consistent with the Dobashi et al. (1994) $^{13}$CO survey of the same
region.

\begin{figure}
\plotfiddle{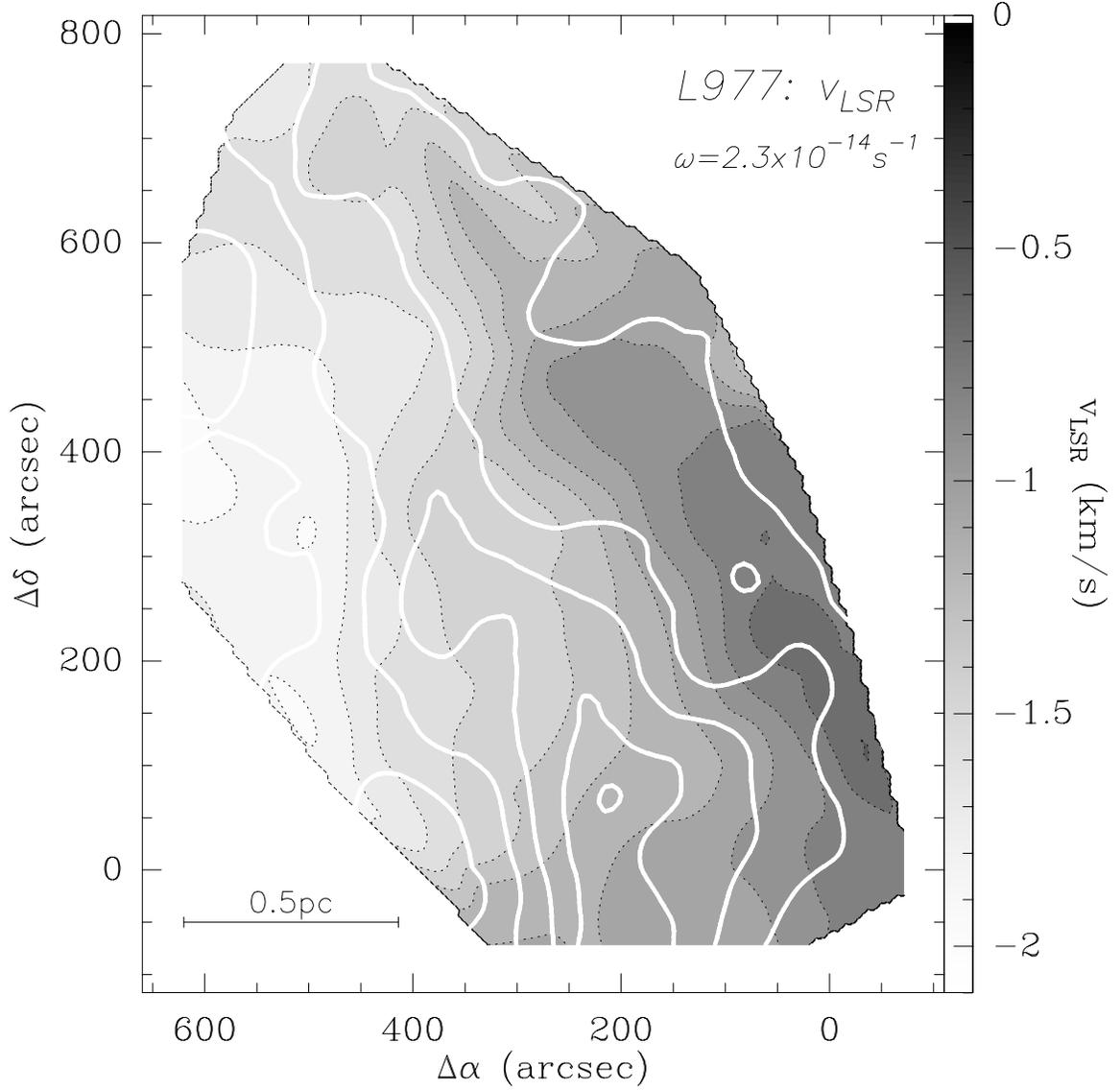}{14cm}{0}{85}{85}{-260}{-60}
\caption[Variation of the LSR velocity across the
surveyed area] {Variation of the LSR velocity across the
surveyed area (grayscale).  The solid white contours represent the
C$^{18}$O integrated intensity.  It is clear from this Figure
that the line velocities show a systematic spatial pattern with a well
defined linear gradient ($\sim 1.2$ kms$^{-1}$ pc$^{-1}$ for an
assumed distance to the cloud of 500pc) along the East--West
direction, essentially orthogonal to the dark filament. \label{vlsr}}
\end{figure}

Figure~\ref{vlsr} shows in grayscale the variation of the LSR velocity
across the surveyed area.  The solid white contours represent the
C$^{18}$O integrated intensity (smoothed to a 100\as\ resolution for the
sake of the clarity of the Figure.  See below for more discussion on the
smoothing procedure).  It is clear from Figure~\ref{vlsr} that the line
velocities show a systematic spatial pattern with a well defined linear
gradient ($\sim 1.2$ kms$^{-1}$ pc$^{-1}$ for an assumed distance to the
cloud of 500pc) along the East--West direction, essentially orthogonal to
the dark filament.  This systematic motion could be either the result of
rotation or shear.  If we assume the filament to be in solid--body
rotation (uniform angular velocity about the axis of symmetry of the
filament) we can derive an angular velocity for this cloud of $\omega =
2.3\times10^{-14}$ s$^{-1}$.

\subsection{A Gaussian Convolved Map of Dust Extinction}

The near--infrared data used in this paper is taken from Alves et al.
(1998).  A detailed description of the near--infrared data acquisition and
data reduction can be found in the latter paper.  The area surveyed in the
3 near--infrared bands ($J$ (1.25 $\mu$m), $H$ (1.65 $\mu$m), and $K$
(2.20 $\mu$m)) towards L977 is presented in Figure 1 of Alves et al. 
(1998) and is similar to the molecular--line survey presented in this
paper (Figure 1). 

The main goal of this paper is to study the variation of molecular
abundance of C$^{18}$O with dust extinction for the L977 molecular cloud. 
While the CO observations are obtained with a radio--telescope with a
diffraction limited Gaussian beam ($\sim 50^{\prime\prime}$ FWHM) the
infrared observations are obtained with an imaging camera and colors and
extinctions are measured along individual, very high resolution
\footnote{The expected angular size of a stellar disk of a field star
behind the cloud is $\la$ $10^{-5} - 10^{-4}$ arcsec.}, pencil--beams
towards individual background stars.  In order to enable a direct
comparison between these two observables we constructed a new extinction
map by spatially convolving the near--infrared measurements with a 2D
Gaussian filter with a FWHM of 50\as\ --- similar to the beam--size of the
FCRAO radio-telescope --- and centered on the positions of the
molecular--line data.  The Gaussian filter was truncated at $r = 3\sigma$,
where $r$ is the distance from the pixel center (the peak of the 2D
Gaussian) and $\sigma$ is its standard deviation.  To improve the signal
to noise ratio we constructed a second set of maps by smoothing both the
dust extinction and the C$^{18}$O measurements to an angular resolution of
100\as.  The smoothing was done in the following manner:  for
molecular--line data, every contiguous 2$\times$2 set of spectra were
averaged into one spectrum that was taken to be at the central position of
the set, equally distant from the original positions of the four spectra
(the outcome of this procedure is a new, Nyquist sampled, map of the
molecular--line data with an effective resolution of 100\as); for the
extinction data, an extinction map with a FWHM resolution of 100\as\ was
then constructed to match the re-gridded molecular--line data. 

\begin{figure}
\plotfiddle{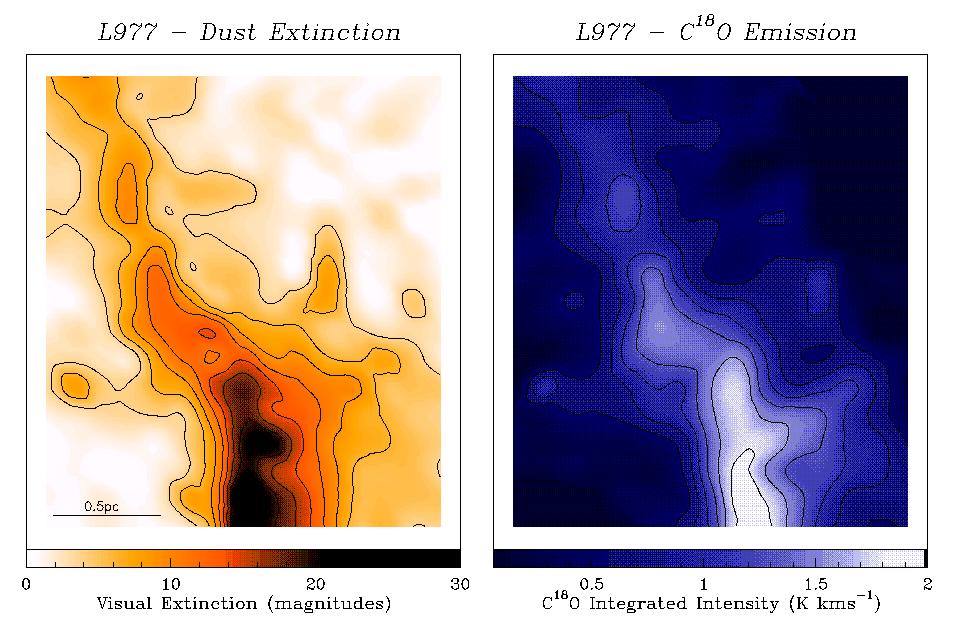}{14cm}{-90}{75}{75}{-290}{410}
\caption[Gaussian convolved map of the dust
extinction and integrated C$^{18}$O molecular--line emission] 
{ a) Gaussian convolved map of the dust
extinction towards the L977 molecular cloud.  Contours start at 5
magnitudes of visual extinction and increase in steps of 2.5
magnitudes.  b) integrated C$^{18}$O molecular--line emission map as
derived from the observations presented in this paper.  Contours start
at 0.5 K kms$^{-1}$ and increase in steps of 0.25 K kms$^{-1}$.  Both
maps are presented at a 100$^{\prime\prime}$ resolution.  
Both maps correlate well not only between themselves but also with
the visually opaque regions of L977 as seen in Figure~\ref{palomar}.
\label{extgasbw}}
\end{figure}

In Figure~\ref{extgasbw} a) we present a Gaussian convolved map of the
dust extinction towards the L977 molecular cloud.  Contours start at 5
magnitudes of visual extinction and increase in steps of 2.5 magnitudes. 
In Figure~\ref{extgasbw} b) we present the integrated C$^{18}$O
molecular--line emission map as derived from the observations presented in
this paper.  Contours start at 0.5 K kms$^{-1}$ and increase in steps of
0.25 K kms$^{-1}$.  Both maps are presented at a 100$^{\prime\prime}$
resolution.  A visual comparison between between these two column density
tracers reveals a notably good correlation.  Also, both maps correlate
well with the visual opaque regions of the L977 cloud as seen in
Figure~\ref{palomar}.

\section{Analysis and Discussion}
\subsection{Comparison Between Gas and Dust in L977}

\begin{figure}
\plotfiddle{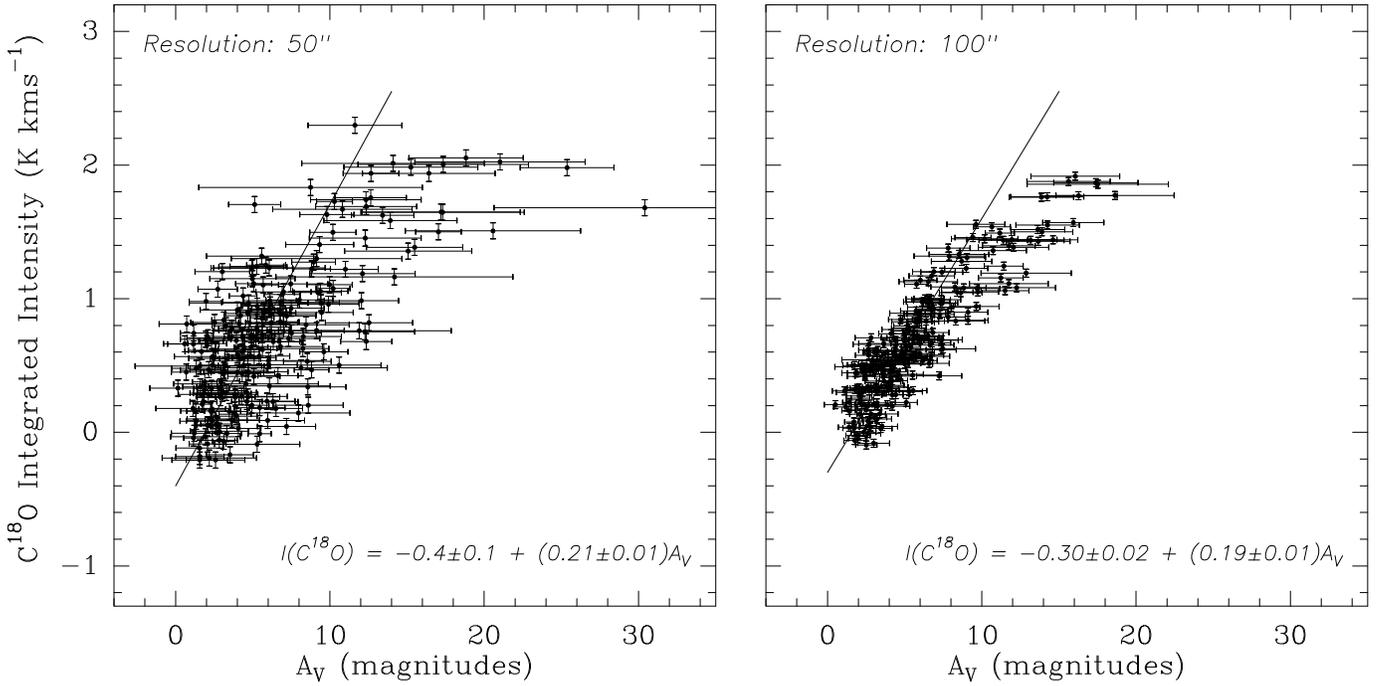}{14cm}{-90}{80}{80}{-310}{430}
\caption[Comparison between visual extinction and
C$^{18}$O integrated intensity for the 240 positions in the L977
molecular cloud] {a) Comparison between visual extinction and
C$^{18}$O integrated intensity for the 240 positions in the L977
molecular cloud.  The error in the extinction measurements were
estimated from the Gaussian weighted rms dispersion of dust column
density measurements falling inside each extinction ``beam'' while the
errors in the integrated intensities were standardly derived from the
molecular--line spectra.  The solid straight line represents the
result of a linear least--squares fit, with errors in both coordinates,
over the entire data set. b) same as a) but smoothed to a resolution
of 100\as. \label{correl}}
\end{figure}

The correlation between visual extinction and C$^{18}$O integrated
intensity is displayed in Figure~\ref{correl} a).  It compares 240
positions where common data were obtained at an angular resolution of
50\as.  The errors in the extinction measurements were estimated from the
Gaussian weighted rms dispersion of dust column density measurements
falling inside each ``beam'' while the errors in the integrated
intensities were directly derived from the noise in molecular--line
spectra.  In Figure~\ref{correl} b) we present the same data as in
Figure~\ref{correl} a) but smoothed to a 100\as\ resolution.  The observed
smaller maximum in the extinction measurements is a consequence of the
smoothing.

At first approximation there is a clear correlation between these two
column density tracers.  A bivariate linear least--squares fit over the
entire data set, represented on Figure~\ref{correl} by the solid straight
line, gives for the two resolutions: 

\begin{equation}
I_{L977}(C^{18}O) \:\, = -0.4 \pm 0.1 +(0.21 \pm 0.01)\,A_V \,\,\:\, 
K\,kms^{-1} \,\,\,\,\,\,\,\, (50^{\prime\prime} \,\,resolution) 
\end{equation}

\begin{equation}
I_{L977}(C^{18}O) \:\, = -0.30 \pm 0.02 +(0.19 \pm 0.01)\,A_V \,\,\:\, 
K\,kms^{-1} \,\,\,\,\,\,\,\, (100^{\prime\prime} \,\,resolution) 
\end{equation}

\noindent which within the uncertainties of the fit, are the same. We note
that the derived slopes for both correlations are affected by the deviant
points and they become somewhat steeper ($\sim$ 10\%) if data above A$_V =
10$ mag are excluded from the bivariate fit. 

However, there is also a notable deviation from this linear correlation
for A$_V$ $\ga$ 10 magnitudes.  There are three effects that could
account for this deviation:  either 1) C$^{18}$O becomes optically thick
at cloud depths A$_V$ $\ga$ 10 magnitudes, 2) there is a systematic
decrease of the excitation temperature down to extremely low values
($T_{ex} < 5$ K at A$_V$ $\ga$ 10 mag), or 3) CO depletion onto dust is
occurring in the high extinction regions of this cloud. The present study
cannot discriminate between these three possible different explanations. 
Still, simple considerations can be used for a preliminary evaluation of
these alternatives: 

\begin{enumerate} 

\item {\it Line saturation:} Observations of CO isotopomers optically
thinner than C$^{18}$O (e.g., C$^{17}$O, $^{13}$C$^{18}$O) are required to
test whether the C$^{18}$O lines are beginning to saturate for A$_V$
$\ga$ 10 magnitudes. Unfortunately such observations are not yet
available and we cannot rule out this possibility. 

\item {\it Extremely low temperature:} If the C$^{18}$O emission were
optically thin, then a possible alternative explanation for the relative
decrease in line intensity at high A$_V$ could be that the C$^{18}$O
excitation temperature at high extinction is so low ($T_{ex} <$ h$\nu$/k
$\sim$ 5 K) that the gas is too cold to emit significantly above the 3 K
background.  We cannot easily rule out this alternative as a viable
explanation for the observed deviation from the linear correlation.
However this would likely require a significant drop in the gas heating
rate at high extinction.  Cosmic rays are believed to be the dominant
heating mechanism of molecular clouds for A$_V > 5$ magnitudes.  In order
for the gas to cool below 5 K, cosmic rays would have to be very
effectively shielded from the cloud's interior. For example, we can
estimate the cosmic ray ionization rate ($\zeta$) required to maintain the
gas at 5 K, by balancing the heating and cooling rates in the cloud. The
cosmic ray heating is given by: 

\begin{equation}
\Gamma_{CR} = \zeta (H_2) \, \Delta Q_{CR} \, n(H_2) 
\end{equation}

\noindent where $\zeta (H_2)$ is the cosmic ray ionization rate of H$_2$,
$\Delta Q_{CR}$ is the energy deposited as heat as a result of this
ionization, and $n(H_2)$ is the molecular hydrogen density.  The cooling
rate for $n(H_2) \sim 10^5$ cm$^{-3}$ can be expressed as (Goldsmith \&
Langer 1978): 

\begin{equation}
\Lambda = 3.8\times10^{-26} \, T^{2.9} \,\,\, (ergs \, cm^{-3} \, 
s^{-1}) 
\end{equation}

\noindent Balancing the two rates, $\Gamma_{CR} = \Lambda$, we derive the
cosmic ray ionization rate necessary to keep the cloud temperature at T
$<$ 5 K to be $\zeta (H_2) < 1.3\times10^{-18}$ s$^{-1}$. 

Theoretical calculations show that cosmic rays are not easily shielded.
They can easily penetrate substantial amounts of molecular cloud material
(A$_V \sim 100$ magnitudes) and maintain ionization rates of $\zeta (H_2)
\sim 10^{-17}$ s$^{-1}$ through out a cloud (Cesarsky \& V\"olk 1978;
L\'eger, Jura, \& Omont 1985) sufficient to maintain cloud temperatures at
$\sim$ 10 K (Williams et al.  1998).  Moreover, if the cosmic ray flux
were to be strongly attenuated in cloud regions of high column density,
the electron density (n$_e$) would decrease and ambipolar diffusion would
be expected to take place providing an important additional source of
heating from the friction between ions and neutrals as the magnetic lines
diffuse out of these dense regions (Mouschovias 1978; Lizano \& Shu 1987).
Finally, in the case of higher densities (n $>$ 10$^8$ cm$^{-3}$) where
cosmic ray heating might have a reduced role it is believed that natural
radioactivity will become an important heating mechanism (Nakano \&
Tademaru 1972).  These considerations suggest that cloud core temperatures
below 5 K are unlikely.

\item {\it CO depletion:} Solid--state CO absorption features have been
observed in the infrared spectra of stars background to dense molecular
clouds suggesting that an appreciable fraction ($\sim 30$\%) of the total
CO abundance may be depleted onto grains inside the cold, high extinction
regions of these clouds (e.g., Whittet \& Duley 1991). Also, chemical
models of the interaction between gas phase molecules and grain surfaces
inside molecular clouds show that significant depletion of molecules
should occur for T $<$ 20 K (Bergin, Langer, \& Goldsmith 1995). The
observed break in the C$^{18}$O versus A$_V$ relation for L977 could be
then the result of depletion at high extinctions. For example, comparison
of C$^{18}$O observations with extinction measurements of the IC5146
molecular cloud (Lada et al.  1994) shows a similar departure from a
linear relation at A$_V$ $\ga$ 10 magnitudes. Recently, Kramer et al. 
(1998) measured both C$^{18}$O and C$^{17}$O lines in the high extinction
regions of IC5146 molecular cloud and found the C$^{18}$O emission to be
optically thin, thus suggesting that the C$^{18}$O abundance is depleted
by factors of 3 to 4 in the high extinction portions of that cloud.  The
similarity between our results and those of IC5146 may indicate that
appreciable depletion is also occurring in L977.  Presently, we favor the
CO depletion alternative as the most likely cause of the break in the
observed correlation.  However, a determination of the opacity of the
C$^{18}$O emission would be required to conclusively test this suggestion.

\end{enumerate}

The derived slope in the C$^{18}$O versus A$_V$ relation for L977
(0.19$\pm$0.01) is nearly twice as steep as the one derived for the IC5146
molecular cloud (0.10$\pm$0.01) from a similar set of observations. 
However, we do not believe that this difference is real.  Two different
aspects of our analysis procedure can explain the observed differences in
slope between these two clouds:  1) we performed a bivariate
least--squares fit (hence we are not as sensitive to the less accurate,
most deviant, measurements at high extinction), and 2) we used a Gaussian
beam to smooth the extinction measurements which allows a better
comparison with the radio observations than the square spatial filter used
by Lada et al.  (1994).  When the IC5146 data is analyzed in the same
manner as the L977 data (the effective resolution of the Lada et al.
(1994) molecular--line data is 102$^{\prime\prime}$), the correlation
between dust extinction and C$^{18}$O integrated intensity is found to be: 

\begin{equation}
I_{IC5146}(C^{18}O) \:\, = -0.30 \pm 0.03 +(0.18 \pm 0.01)\,A_V \,\,\:\, 
K\,kms^{-1} \,\,\,\,\,\,\,\,  
\end{equation}

\noindent which within the uncertainties is indistinguishable from the one
derived for L977 cloud.  The difference between the earlier Lada et al.
fit and that presented here is almost entirely a result of the use of a
Gaussian spatial filter which was well matched to the radio beam.
Moreover, the improved fit is in better agreement with the higher
resolution C$^{18}$O observations of Kramer et al. (1998) obtained with
the IRAM 30m telescope. 

In the present study, we derive a negative intercept in the relations for
both clouds which implies that the survival of the C$^{18}$O molecule in
the ISM requires a shielding from the interstellar radiation field (ISRF)
equal to a dust column density equivalent of 1.6$\pm$0.5 magnitudes of
visual extinction.

\subsection{Molecular Abundance of C$^{18}$O}

Our extinction measurements toward the L977 molecular cloud allow us to
directly determine the abundance of C$^{18}$O over a broad range of cloud
depths.  To do this we need to convert the C$^{18}$O integrated
intensities to column densities at each position in the map and compare
the derived column densities with the extinction measurements derived from
the near--infrared data.  Because we have single--transition
molecular--line data (C$^{18}$O (1--0)) we can only derive LTE column
densities for this molecule.  This is an approximation and we expect
non--LTE effects to be operating in this cloud which in turn will cause
our derived column densities to be somewhat inaccurate.  Still, column
densities determined in this way can be compared to similar existing
studies and even to future studies that make use of this standard
approximation. 

\begin{figure}
\plotfiddle{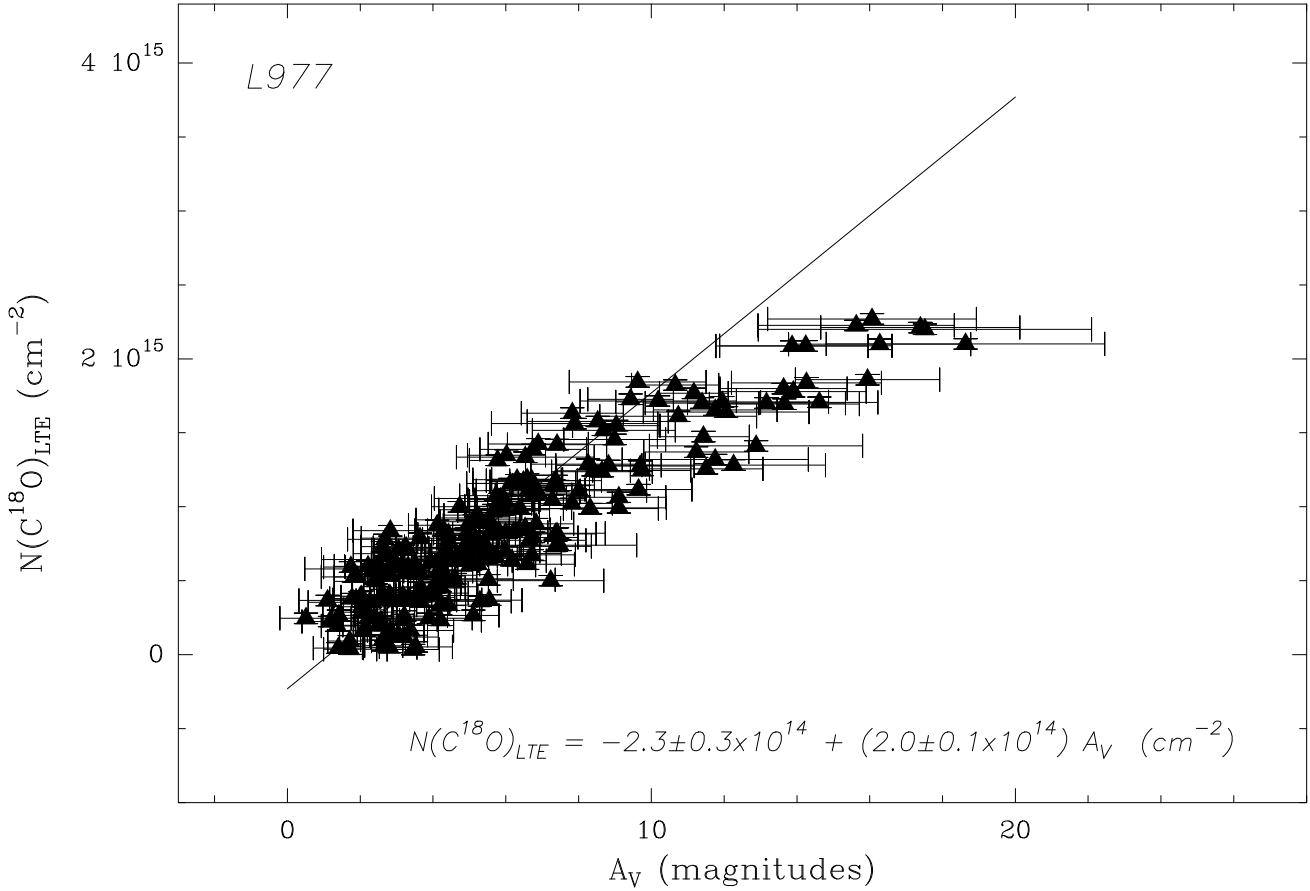}{14cm}{-90}{75}{75}{-290}{430}
\caption[Relation between N(C$^{18}$O)$_{LTE}$ and visual
extinction A$_V$] {Relation between N(C$^{18}$O)$_{LTE}$ and visual
extinction A$_V$. The errors in N(C$^{18}$O)$_{LTE}$ are smaller than 
the size of the symbols. The solid straight line represents the
result of a linear least--squares fit, with errors in both coordinates,
over the entire data set. There is a clear deviation from the linear relation
at high extinctions. \label{abun}}
\end{figure}

LTE column densities were derived for each point of the molecular--line
grid assuming a C$^{18}$O excitation temperature of 10 K (Dobashi et al.
1994).  In Figure~\ref{abun} we plot the derived abundances as function of
cloud depth for our 100\as\ resolution data.  The errors in
N(C$^{18}$O)$_{LTE}$ are smaller than the size of the symbols.  In order
to quantify the observed correlation we performed a least--squares fit
with errors on both coordinates.  The derived fit, shown as a straight
line in Figure~\ref{abun}, is:

\begin{equation}
N_{L977}(C^{18}O)_{LTE} \:\, = -2.3 \pm 0.2\times 10^{14} + 
(2.0 \pm 0.1\times 10^{14})\,A_V \,\,\:\, 
cm^{-2}  
\end{equation}

The same fit performed on the re--analyzed IC5146 data returns:

\begin{equation}
N_{IC5146}(C^{18}O)_{LTE} \:\, = -2.5 \pm 0.2\times 10^{14} + 
(2.1 \pm 0.1\times 10^{14})\,A_V \,\,\:\, 
cm^{-2}  
\end{equation}

\noindent which once again is indistinguishable from the fit done to the
L977 data.  These results are also compatible with the range of values
derived for comparable optical depths by Frerking et al.  (1982).  These
authors performed pointed radio observations toward known reddened single
background stars in Taurus and $\rho$ Oph molecular cloud complexes and
found the slope of the N(C$^{18}$O) vs. A$_V$ relation to be 2.4 for
Taurus and 1.7 for $\rho$ Oph (uncertainties not quoted).

\subsection{How Much Mass is C$^{18}$O Missing?}

In section 4.1 of this paper we concluded that there is a break in the
correlation between dust extinction and the C$^{18}$O integrated intensity
at large cloud depths.  This could be due to C$^{18}$O opacity effects,
extremely low ($T_{ex} <$ 5 K) excitation temperatures, or, more likely
(see Section 4.1), to the depletion of this molecule onto the dust grains
in the denser regions of L977.  In any event, the observed break implies
that above a certain threshold of cloud depth C$^{18}$O cannot be taken as
a reliable tracer of molecular cloud mass.  Since C$^{18}$O emission is
widely used as good tracer of column density in molecular clouds it is
important to account for this limitation and to make an estimation of its
magnitude. 

\begin{figure}
\plotfiddle{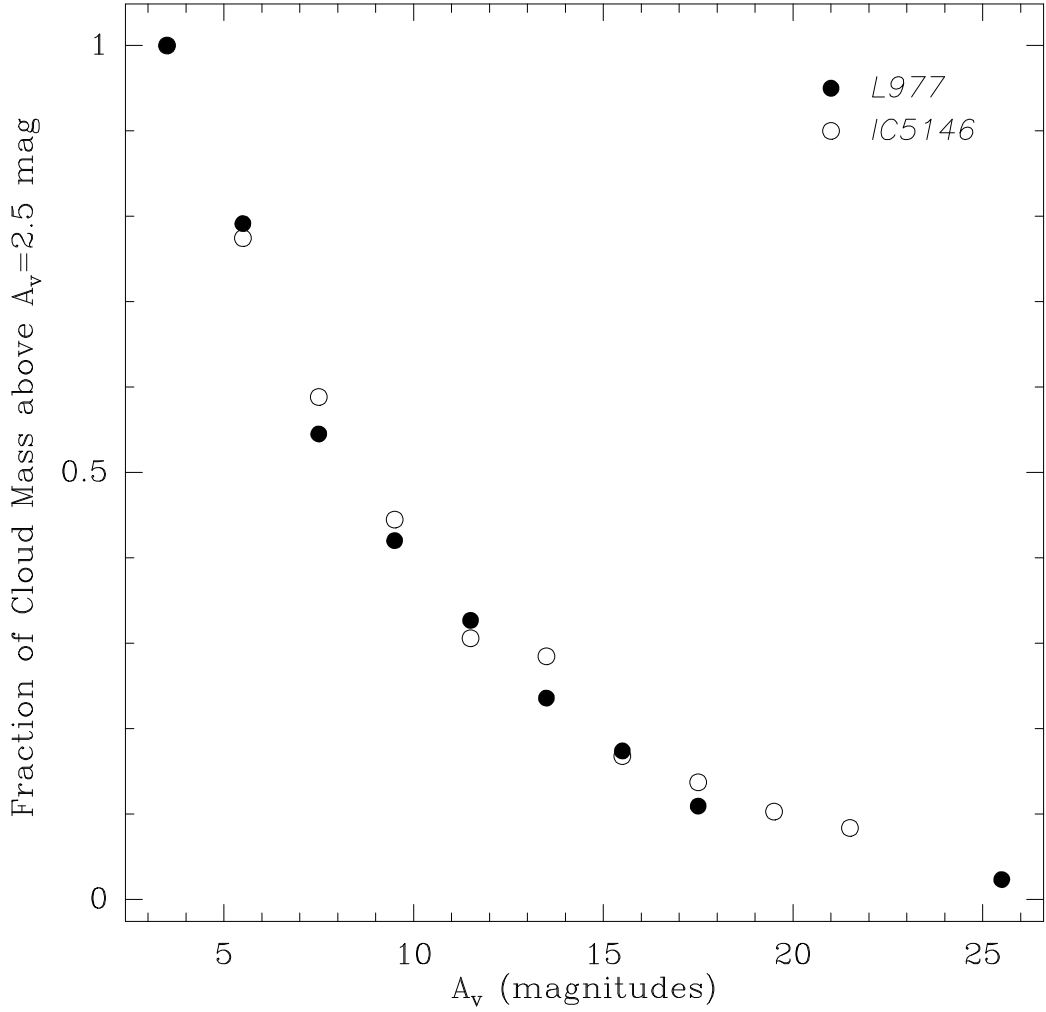}{14cm}{-90}{90}{90}{-350}{500}
\caption[Mass distribution in L977 and IC5146 molecular cloud
as a function of cloud depth] {Mass distribution in L977 and IC5146 
molecular cloud
as a function of cloud depth. There is a remarkable similarity
between the two mass distributions  
suggesting a similar physical structure for these two clouds. 
\label{massfrac}}
\end{figure}

Figure~\ref{massfrac} shows the distribution of cloud mass derived from
the near--infrared extinction measurements as a function of cloud depth
for both L977 and IC5146 molecular clouds.  The data points are the
cumulative binning, in intervals of 2 magnitudes of visual extinction, of
the normalized cloud masses contained in the extinction map pixels with
average extinctions above 2.5 magnitudes. There is a remarkable similarity
between the two molecular clouds (Alves et al.  1998).  Note that the
results in Figure~\ref{massfrac} are independent of distance.  Since in
the most dense regions of both clouds we cannot measure extinction --- our
observations are not sensitive enough to detect heavily reddened (A$_V >$
30 mag) stars --- this plot is not complete at the high extinction end. 
Nevertheless, the solid angles occupied by these regions are very small
when compared to the total solid angle of the cloud.  So, although we are
missing the densest regions in each cloud their contribution to the
overall distribution of mass is negligible.

From Figure~\ref{correl} we concluded that a break in the correlation
between dust and C$^{18}$O integrated intensity occurs as early as cloud
depths of about 10 magnitudes of visual extinction and that this same
correlation appears to flatten out for cloud depths \gtsima 20 magnitudes. 
According to the distribution of molecular cloud masses displayed in
Figure~\ref{massfrac} for clouds L977 and IC5146 this fact implies that
when using C$^{18}$O integrated intensity as a column density tracer we
may be missing between 10 to 30\% of a cloud's mass.  In the case of L977
and for A$_V >$ 2.5 mag, the mass as traced by C$^{18}$O is $\sim$ 15\%
smaller than that traced by dust.  Nevertheless, this systematic error in
the determination of L977 mass or other molecular clouds masses is still
small when compared to the systematic uncertainty introduced in the same
determination by the poorly known distances to these objects (typically a
factor of 2).  However, the systematic error introduced by using C$^{18}$O
emission to trace mass is a function of cloud depth.  For example,
C$^{18}$O would be a very poor mass tracer of dense (A$_V > 10$) cores
inside a molecular cloud.

\section{Summary}

We studied the correlation between $J =$ (1--0) C$^{18}$O emission and
dust extinction for the L977 molecular cloud over a large range of cloud
depth (2 $<$ A$_V$ $<$ 30 mag).  Our main findings are as follow: 

\begin{itemize}

\item{There is a good linear correlation between these two column density
tracers for cloud depths A$_V$ $\la$ 10 magnitudes; for cloud depths
above this threshold there is a notable break in the correlation of the
two mass tracers.}

\item{Although either optically thick C$^{18}$O emission or extremely low
($T_{ex} <$ 5 K) excitation temperatures could produce this break, CO
depletion in the denser, coldest regions of L977 may be the most likely
cause of the departure from linearity at high extinctions in the observed
correlation.  Observations of rarer CO isotopomers are needed to
distinguish between these alternatives.}

\item{Both the C$^{18}$O abundance and its variation over a broad range of
cloud depths (A$_V$ $\la$ 10 magnitudes) is directly derived and found
to be virtually the same as that derived for the IC 5146 molecular cloud
in a similar study by Lada et al (1994). It is also in agreement with
previous determinations in other clouds.}

\item{In regions of very high extinction (A$_V > $ 10 mag), such as dense
molecular cores, our results suggest that C$^{18}$O would be a very poor
tracer of mass. Consequently, using C$^{18}$O as a column density tracer
in molecular clouds can lead to a 10 to 30\% underestimation of overall
cloud mass.}

\item{We estimate the minimum shielding required for the survival of the
C$^{18}$O from the interstellar radiation field to be 1.6 $\pm$ 0.5
magnitudes of visual extinction.}

\end{itemize}

\acknowledgments

It is a pleasure to acknowledge Mario Tafalla for discussions and for
assistance on the reduction of the molecular--line data.  We acknowledge
Ted Bergin and Jonathan Williams for fruitful discussions. We also
acknowledge an anonymous referee for pointing out extremely low
temperature gas as a possible explanation to our results. This study was
supported by the Smithsonian Institution Scholarly Studies Program
SS218--3--95.  J. Alves acknowledges support from the Funda\cao\ para a
Ci\^encia e Tecnologia (FCT) Programa Praxis XXI, graduate fellowship
BD/3896/94, Portugal.

\end{document}